\documentclass[10pt, a4paper]{IEEEtran}

\usepackage{amsmath,amssymb,amsthm}
\usepackage{algorithmic}
\usepackage{array}
\usepackage{fixltx2e}
\usepackage{stfloats}
\usepackage[utf8]{inputenc}	
\usepackage[british]{babel}
\usepackage{csquotes}
\usepackage[pdftex]{graphicx}
\graphicspath{ {image/} }
\usepackage{amssymb}
\usepackage{mathtools}
\usepackage{todonotes}
\usepackage{authblk}

\begin{document}
\title{Time Reversal Precoding at SubTHz Frequencies: Experimental Results on Spatiotemporal Focusing}

\author[1]{Ali Mokh}
\author[1]{Julien de Rosny}
\author[2]{George C. Alexandropoulos}
\author[3]{\\Ramin Khayatzadeh}
\author[1]{Abdelwaheb Ourir}
\author[3]{Mohamed Kamoun}
\author[1]{Arnaud Tourin}
\author[1]{Mathias Fink}

\affil[1]{ESPCI Paris, PSL Research University, CNRS, Institut Langevin, France}
\affil[2]{Department of Informatics and Telecommunications,
National and Kapodistrian University of Athens, Greece}
\affil[3]{Mathematical and Algorithmic Science Lab, Paris Research Center, Huawei Technologies France}
\affil[ ]{emails: firstname.lastname@espci.fr, alexandg@di.uoa.gr, firstname.lastname@huawei.com}



\maketitle

\begin{abstract}
Due to availability of large spectrum chunks, the sub-TeraHertz (subTHz) frequency band can support Ultra-WideBand (UWB) wireless communications, paving the way for unprecedented increase in the wireless network capacity. This fact is expected to be the next breakthrough for the upcoming sixth Generation (6G) standards. However, the technology of subTHz transceivers is not yet mature enough to apply the advanced signal processing currently being implemented for millimeter wave wireless communications. In this paper, we consider the Time Reversal (TR) precoding technique, which provides simple and robust processing capable to offer highly focalized in time and space UWB waveforms, exploiting the spatial diversity of wireless channels.    
We first investigate experimentally the performance of subTHz TR focusing in complex media inside a leaking reverberation
cavity. We then combine TR with received spatial modulation to realize data communication using a simple non-coherent receiver with two antennas. Our results showcase the capability of TR to offer focusing in time in the order of few nanoseconds and in space in the order of less than $1$ mm.
\end{abstract}

\begin{IEEEkeywords}
Time reversal, receive spatial modulation, terahertz communications, spatiotemporal focusing.
\end{IEEEkeywords}

\IEEEpeerreviewmaketitle
\section{Introduction}
The last decade has witnessed a tremendous growth in the global mobile data traffic, and by 2030, the trend of exponential data rate increase will continue, overcoming the capabilities of the fifth Generation (5G) networks. To confront the shortcomings of 5G, initial requirements for the future 6G communication networks are being discussed dealing with wider frequency bands as well as the needs for increasing the maximum transmission rate, better spectrum efficiency, extremely massive network connections and antenna deployments, shorter delays, increased coverage, and improving anti-interference capability \cite{zhao2021survey,shlezinger2020dynamic}.
The initial discussions indicate that the frequencies in the range of Sub-TeraHertz (SubTHz) and above are highly probable to be considered for 6G, since there are plenty of available bands suitable to satisfy those early demanding requirements \cite{saad2019vision,tariq2020speculative}. In fact, it appears that 6G will utilize spectrum beyond $140$ GHz with particular applications in very short-range communication, or the so-called \textit{whisper radio} \cite{xing2018propagation}.
However, because of the current technological limits, it is difficult to apply in SubTHz the signal processing currently in use for millimeter wave communications, such as Orthogonal Frequency-Division Multiplexing (OFDM). In this paper, we consider the Time Reversal (TR) technique \cite{fink1999time} as the pre-processor for robust data transmissions, which can be paired with a simple receiver implementing basic reception signal processing. 

A TR mirror is an apparatus that is capable of focusing waves both in time and in space, i.e., of offering spatiotemporal focusing. It is based on the time-reversal symmetry of propagating media. Thirty years ago, it has been first applied to focus ultrasonic waves in water, tissues, and solids. The TR mirror is very efficient in complex media such as multiple scattering media \cite{Derode_1995} or cavities\cite{Draeger_1997}, because it takes benefit of wave propagation reflections to enhance its spatiotemporal focusing. In acoustics, the main application of TR has been non-destructive control, imaging, and therapy \cite{fink1999time}. In 2004, the concept of the TR mirror was extended to electromagnetic waves \cite{lerosey2004time} realized at $2.4$ GHz. Two years later, capitalizing on the fact that the larger the transmission bandwidth the better the focusing is, wideband TR was successfully realized \cite{lerosey2006time}. In \cite{Qiu_JSAC_2006}, TR was considered for ultra-wideband communications. The improvement in the Signal-to-Noise (SNR) ratio, thanks to the TR-based temporal focusing, has been experimentally verified with a multi-antenna transmitter in \cite{TR_MISO_AWPL_2006, mokh2021indoor}. The concept of TR division multiple access was introduced in \cite{TR_MIMO_MA_2012} for multi-user communication. Very recently, \cite{Alexandropoulos_ICASSP} demonstrated indoor cm-level localization accuracy with TR at $3.5$ GHz using up to $600$MHz bandwidth channel sounding signals.
 
In this paper, we experimentally study the spatiotemporal focusing properties of time-reversed electromagnetic waves centered at $273.6$ GHz with a bandwidth up to $4$ GHz in multipath complex media realized with a leaking reverberation cavity. The experimental measurements are followed by a simulation evaluation intending to verify the possibility of data transmission towards a simple non-coherent receiver equipped with two antenna elements. In particular, we deploy a Receive Spatial Modulation (RSM) chain based on a simple power detector to realize wireless communications at the considered subTHz frequency band.

The rest of the paper is organized as follows. In Section II, the principle of TR-based precoding is described. In Section~III, we present our subTHz experimental setup to realize and measure TR-precoded signals. In Section IV, we present our simulated evaluation of the Bit Error Rate (BER) performance of a communication scheme based on RSM using the measured signals. Finally, the concluding remarks of the paper are included in Section~V.

\section{Time Reversal Precoding}
Let us consider a multiple access wireless communication system comprising one single-antenna transmitter that wishes to communicate in the downlink direction with $N$ single-antenna receivers. We represent by $h_{i}[k]$ the baseband Channel Impulse Response (CIR) at discrete time $k$ between the transmitter and the $i$-th ($i=1,2,\ldots,N$) receiving antenna. The TR precoding utilizes the time reversed CIR, i.e., $h^*_{i}[L-k]$ with $L$ denoting the number of the significant channel taps, to focus the information-bearing electromagnetic field on the $i$-th receiving antenna. In mathematical terms, the TR-precoded signal sent by the transmit antenna in order to focus each baseband information message $x_i[l]$ on its respective receiving antenna is given by the following expression:
\begin{align}
     s[k]=\sum_{i=1}^N  \frac{\sum_{l=1}^L x_i[l] h_{i}^*[L+l-k]}{\sqrt{\sum_{l=1}^L \lvert h_{i}[l] \rvert^2}}. 
     \label{eq:yi}
\end{align}
The normalization factor ensures that the power emitted toward each single-antenna receiver is the same. By using the latter expression, the baseband equivalent of the signal at each $j$-th ($j=1,2,\ldots,N$) receiving antenna can be expressed as
\begin{equation}
   y_j[k]= \sum_{l=1}^L \sum_{i=1}^N x_i[l] R_{j,i}[L+l-k]+n_j[k],
   \label{eq:TR}
\end{equation}
where $R_{j,i}[k]$ represents the correlation function between the CIRs of the $j$-th and $i$-th receiving antennas, which is given as follows: 
\begin{equation}
 	R_{j,i}[k] = \frac{\sum_{k'} h_{i}^*[k-k'] h_{j}[k']}{\sqrt{\sum_{l=1}^L \lvert h_{i}[l] \rvert^2}}.
   \label{eq1}
\end{equation}
In the latter expression, $n_j[\cdot]$ denotes the zero-mean additive white Gaussian noise at the $j$-th receiver having standard deviation $\sigma$. In multipath wireless communication channels, when applying TR precoding with the CIR toward the $i$-th receiving antenna, the time-reversed field focuses in time and in space at this antennas, exhibiting a signal peak. The amplitude of this peak is given by the CIR's autocorrelation function at lag $0$, i.e., by $R_{i,i}[0]$. 
\section{SubTHz Experimental Measurements}
In this section, we present the experimental setup used to investigate the TR spatiotemporal focusing in the considered subTHz frequency band. In such frequencies, the attenuation of the channel becomes very high, and it is difficult to realize a multipath channel even for indoor application. Since such a channel is prerequisite for TR, we have designed a specific setup for each experiment that results in rich scattering fading conditions. Finally, we present the experimental results for the spatiotemporal focusing capability of the considered TR-based precoding scheme.

\subsection{Experiment Setup}
\begin{figure}
\centering
\includegraphics[width=0.7 \linewidth]{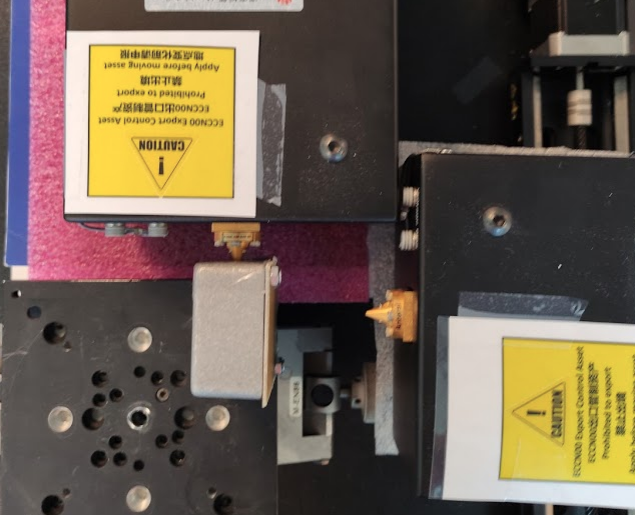}
\includegraphics[width=0.7 \linewidth]{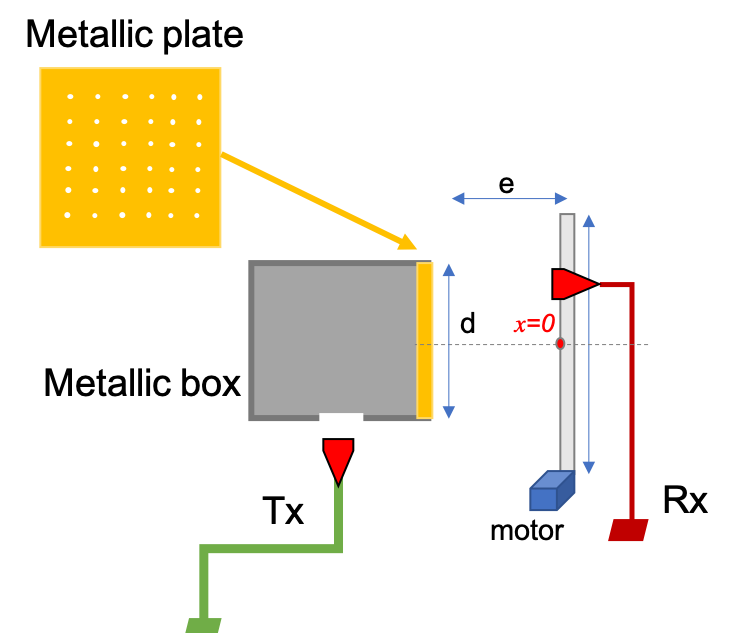}
\caption{Photo and schematic illustration of the designed experimental setup centered at the subTHz frequency $273.6$ GHz with up to $B=4$ GHz bandwidth.}
\label{setup}
\end{figure}
Our experimental subTHz setup operating at the central frequency $f_{c}=273.6$ GHz using a 5 cm $\times$ 5 cm $\times$ 2.5 cm closed metallic box is depicted in Fig$.$~\ref{setup}. The box behaves as a rich multipath medium. To emit inside the cavity, the Transmit (Tx) antenna is inserted into the box through a small hole. To be able to focus outside the box, there are multiple small holes with a diameter of $1$ mm on the other side of the box. The Receive (Rx) antenna is positioned in front of this side of the box at a distance of $5$ cm on a motorized axis, which is parallel to the face (see Fig$.$~\ref{setup}). Both the Tx and Rx antennas are identical. They operate between $260$ and $400$ GHz and are omnidirectional. The point $x=0$ on the axis faces the box's center. 

To create transmitted signals at subTHz, we have generated a reference sinusoidal signal at the desired frequency $f_{c}$, as follows. The deployed local oscillator R\&S SGMA produces a sinusoid at $5.7$ GHz, which was then up-converted for transmission using a WR2.2 VDI SAX WM570 harmonic mixer (operating at $260$-$400$ GHz with $-10$ dBm transmit power) with a multiplication factor set to $48$ (note that $5.7\times48=273.6$). At the Rx side, a harmonic mixer with the same multiplication factor was deployed for down-conversion back to $5.7$ GHz. Both the Arbitrary Waveform Generator (AWG) to produce the $3$ GHz baseband signal and the digital sampling oscilloscope to record the field were connected to the same reference clock working at $10$ MHz. The estimation of the CIR and the TR precoding were performed, as follows. First, a chirp signal spanning the range [$f_c-B/2$ $f_c+B/2$] was transmitted by the Tx with $B$ denoting the transmission bandwidth. Synchronously, the signal was probed by the Rx antenna. A desktop computer performed the CIR estimation by correlating the recorded signal with the chirp signal. Even if TR generated high amplitude pulses, the gain was not sufficient to directly probe the pulse with the Rx antenna. For this reason before emission, the CIR flipped in time was also convoluted with the same chirp signal. Similarly to the CIR estimation, the TR-focused signal was reconstructed in the computer using cross correlation.

\subsection{Performance Results}
\begin{figure}
\centering
\includegraphics[width=1 \linewidth]{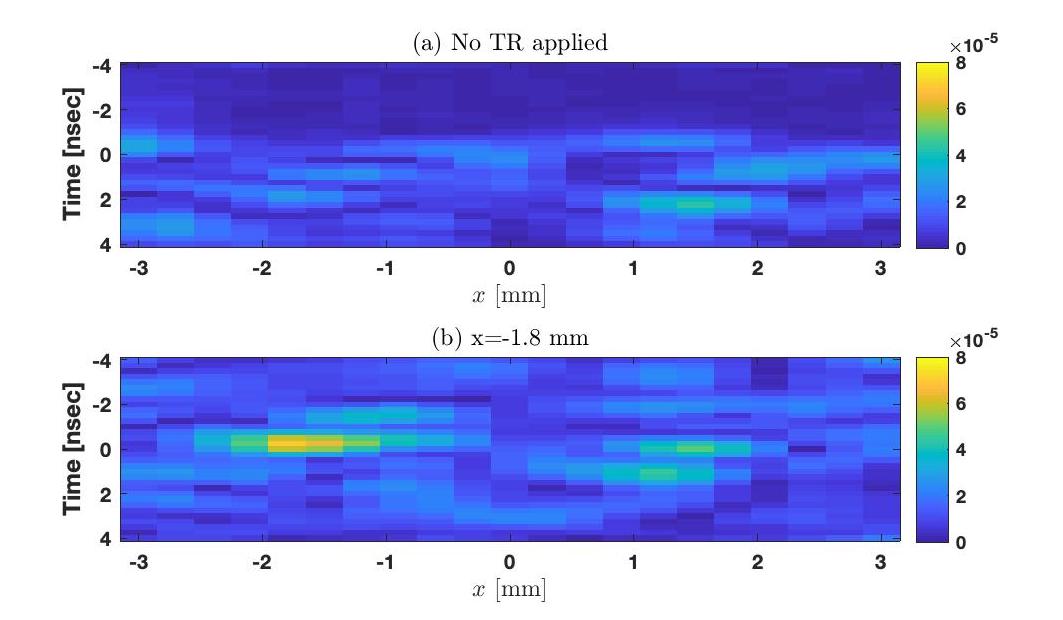}
\caption{(a) Baseband modulus of the CIRs probed at different positions; (b) Baseband modulus of the field probed at different positions when focusing by TR toward the position $x=-1.8$ mm.}
\label{withwithoutTR}
\end{figure}

We have measured the CIR for different positions of the Rx antenna spaced by $0.3$ mm, in particular, between the positions $x=-6.2$ mm and $6.2$ mm, where $x=0$ is the position facing the center of the metallic box. We have used a transmission bandwidth of $B=4$ GHz. Figure~\ref{withwithoutTR}(a) demonstrates the CIRs for those different Rx positions. In Fig$.$~\ref{withwithoutTR}(b), the amplitude of the electromagnetic field is plotted versus the Rx antenna placement when the CIR probed at position $x=-1.8$ mm is flipped in time and emitted. Both figures show the capability of the TR technique to focus the signal in time and space at the targeted Rx position.

\begin{figure}
\centering
\includegraphics[width=1 \linewidth]{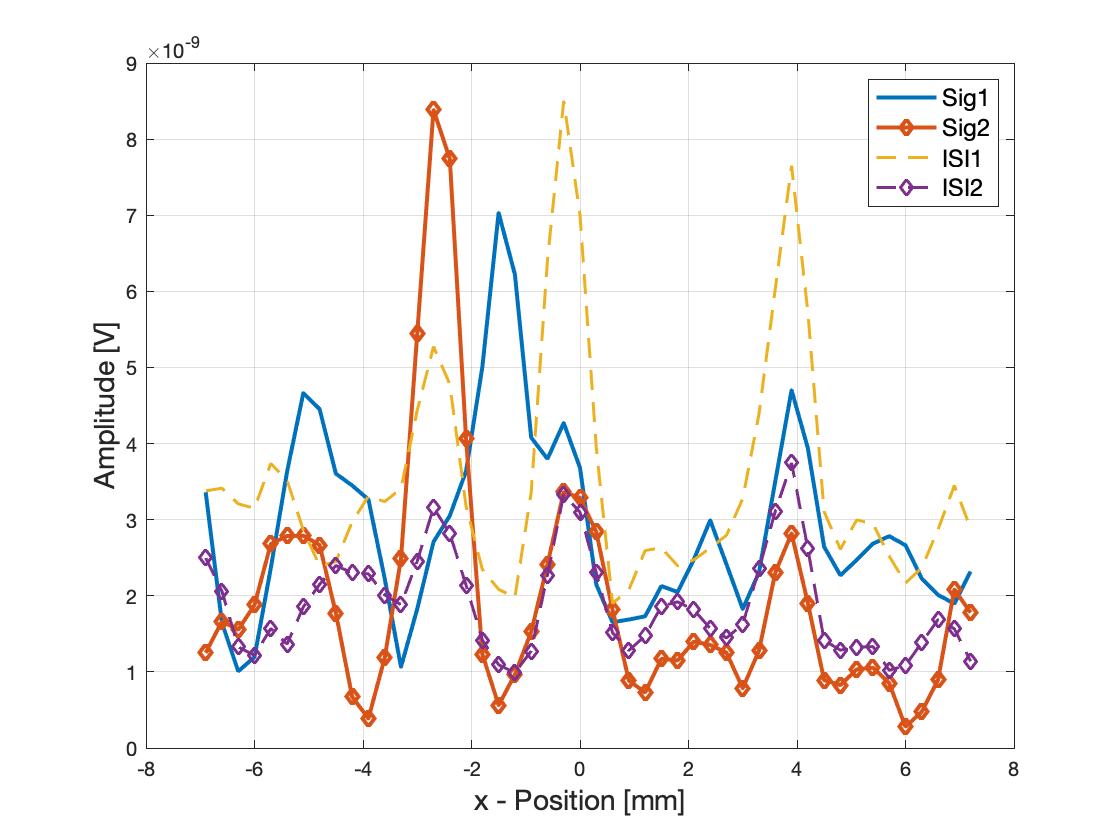}
\caption{The amplitude of the received signal strength and interference from each TR-precoded signal over all possible Rx positions for $D=15$.}
\label{SINR}
\end{figure}

To investigate the TR capability for multiple access scenarios, i.e., spatiotemporal focusing of different signals towards different spatial positions, we applied the TR precoding considering the two Rx positions $x_1=-2.7$ mm and $x_2=-1.8$ mm. The signals have been normalized in such a way that the $2$ received powers are identical. To transmit data, the pulses were modulated either in amplitude or phase. Each pulse was separated by $D$ channel taps. For this case, the duration of each symbol is actually $T_s=\frac{D}{B}$.
From the probing of the $2$ TR-precoded field with respect to the intended Rx positions, we deduced the maximum received signal strengths. Note that the signal strength at the position of the unintended RX antenna provides the Inter-User Interference (IUI). The Inter-Symbol Interferences (ISI) is deduced from the analysis of the temporal secondary lobes. The result for the received signal strengths are included in Fig$.$~\ref{SINR} considering $D=15$. It can be concluded that, for multiple access transmission, the focusing signal at each Rx position is affected by the sum of all interference, namely, teh IUI of the other focusing signal and the two ISI contributions. The resulted signals and corresponding interference will be used in the following section for computing the BER of a multiple access transmission scheme.

\section{Receive Spatial Modulation with TR}
In this section, we make use of the measured subTHz signals to simulate a transmission scheme between the single-antenna transmitter and a receiver with two receive antenna elements. The scheme implements TR-based precoding and an RSM technique based on Receive Antenna Shift Keying (RASK), which targets a specific Rx antenna to transmit data.

\subsection{System Description}
\begin{figure*}
\centering
\includegraphics[width=1 \linewidth]{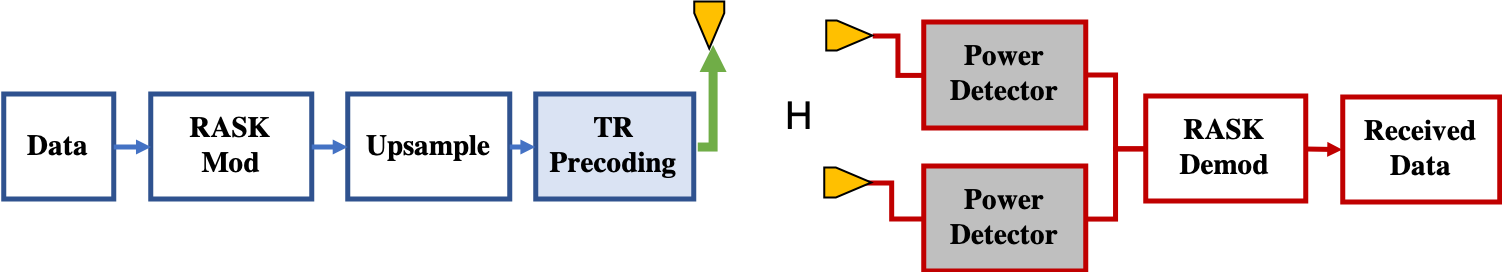}
\caption{Block diagram of the proposed transmission scheme between a single-antenna Tx and an Rx with two antenna elements, which adopts TR precoding and RASK modulation/demodulation with simple power detection.}
\label{BD}
\end{figure*}

The idea is to use TR precoding to target one of two Rx antenna elements \cite{mokh2017time}: target the first antenna to transmit a bit '$0$' and the second antenna to transmit an '$1$' bit. On the Rx side, the receiver needs to understand which antenna has been targeted, and this can be realized with a simple power detector \cite{mokh2018theoretical}. Depending on the estimation of the index of the antenna with the largest received signal strength, one can estimate the transmitted symbol. The block diagram of this simple communication scheme is illustrated in Fig$.$~\ref{BD}. It is noted that a variant of the RASK scheme was proposed in \cite{mokh2017extended} for increasing the spectral efficiency. This scheme was termed as Extended RASK (ERASK), where more combinations can be used, and hence, each Rx antenna can be targeted or not. The only detection requirement here is to estimate if each Rx antenna has been targeted or not, which requires a certain threshold for the received signal strength.

The motivation to use the latter RSM-based schemes lies on our intention to verify the capability of TR to offer spatial multiplexing in the subTHz frequency band, while adopting a very simple receiver that does not need phase detection or synchronization between each receiving antennas.

\subsection{Simulation Results}
\begin{figure}
\centering
\includegraphics[width=1 \linewidth]{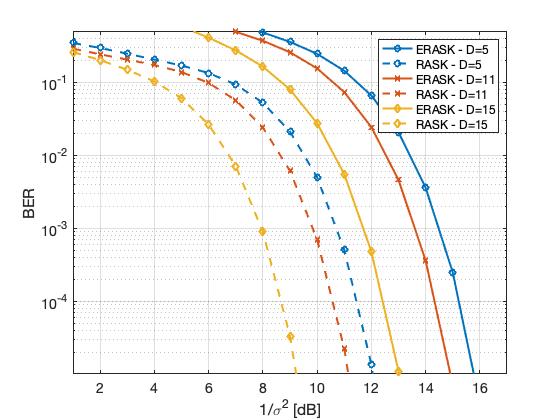}
\caption{BER of RASK and ERASK transmission in subTHz domain towards locations $x=-2.7$ and -1.8 mm, with different $D$}
\label{BER}
\end{figure}

We have used the signals from Section~III with the experimental measurements, and calculated the received signal strength and the interference at the two Rx positions $x_1=-2.7$ mm and $x_2=-1.8$ mm, considering different values for $D$.
Figure~\ref{BER} depicts the corresponding BER when using the RASK and ERASK schemes. Note that the ERASK scheme has a double spectral efficiency ($2$ bits are sent per symbol, while only one bit is sent in the case of RASK). As expected, the greater $D$ is, the better the performance of both two schemes, with the effect of reducing the error performance with increasing SNR $\sigma^{-2}$. These results showcase the capability of our designed TR-based scheme in subTHz frequencies for multiple access applications, offering the very low spatial separation distance of $0.9$ mm.

\section{Conclusion}
In this paper, we experimentally investigated the performance of TR precoding in the subTHz frequency band. An experimental setup has been designed to estimate the CIR at the central carrier frequency $273.6$ GHz with a $4$ GHz bandwidth, which was then used to realize TR-based spatiotemporal focusing. Our measured signals were also used in a simulated data communication scenario including RASK-based reception with simple power detectors. The performance evaluation results showcased the capability of TR to focus transmitted signals in both space and time, thanks to the SubTHz signal wavelength. In addition, our simulation results verified the TR capability for realizing multiple access transmissions towards two receive antennas separated by $0.9$ mm. For future work, we intend to experiment video communication at subTHz considering reconfigurable intelligent surfaces \cite{alexandg_2021} for reprogrammable rich scattering conditions.

\bibliographystyle{IEEEtran}
\bibliography{IEEEabrv,ref}


\end{document}